# Gupta-Rajpoot Texture 2(4) Zero Mass Matrices

Subhash RAJPOOT[1)]

*Department of Physics & Astronomy*
*California State University*
*1250 Bellflower Boulevard*
*Long Beach, CA 90840*

**Abstract**

The so called "Texture 2(4) zero mass matrices" were first introduced in a Wayne State University preprint to accommodate the "yet to be discovered" heavy top quark. The preprint in its entirety is reproduced here after a brief commentary.

[1)] E-Mail: rajpoot@csulb.edu

## Introduction

As the last centuary plunged into its last decade, the two pressing issues confronting the standard model were

1. The top quark
2. The elusive Higgs

Today, the top quark is no longer an issue, but the elusive Higgs still is. Hopefully, LHC will settle the issue in the affirmative.

Concerning the top quark, CDF collaboration in 1989 published their latest result on the lower bound on the mass of the top quark as

$$m_t \geq 89\,\mathrm{GeV} \quad . \tag{1}$$

This being the case, Gupta[2)] and the author undertook a study of many schemes of quark mass matrices and flavor mixing. The most popular scheme at the time was the so-called Fritzsch scheme of quark mixing. In the light of the latest CDF result then, the scheme faced a severe problem, namely the prediction of the top quark mass in the Fritzsch scheme had already hit the CDF lower bound. This bound happened to be almost the upper bound in the Fritzsch scheme. This prompted us to invent new quark mass matrices and examine their implications, should the top quark mass turn out to be much heavier than the CDF lower bound of 89 GeV.

## The Gupta-Rajpoot Ansatz

We found that the scheme [1] that could most elegantly and comfortably accommodate a heavy top consisted of three by three quark mass matrices with nearest neighbour interactions and only two non-vanishing elements in the leading diagonal of the matrices arranged in ascending order in magnitude with the one-one component set equal to zero. Explicitly,

$$M_u = \begin{pmatrix} 0 & A_u e^{i\alpha_u} & 0 \\ A_u e^{-i\alpha_u} & D_u & B_u e^{i\beta_u} \\ 0 & B_u e^{-i\beta_u} & C_u \end{pmatrix} \qquad M_d = \begin{pmatrix} 0 & A_d e^{i\alpha_d} & 0 \\ A_d e^{-i\alpha_d} & D_d & B_d e^{i\beta_d} \\ 0 & B_d e^{-i\beta_d} & C_d \end{pmatrix} \quad . \tag{2}$$

As is well known, the matrices $M_u$ and $M_d$ are usually expressed in the generic form $M = P\mathbf{M}P^\dagger$ where

$$\mathbf{M} = \begin{pmatrix} 0 & A & 0 \\ A & D & B \\ 0 & B & C \end{pmatrix} \quad , \qquad P = \begin{pmatrix} 1 & 0 & 0 \\ 0 & e^{-i\alpha} & 0 \\ 0 & 0 & e^{-i(\alpha+\beta)} \end{pmatrix} \quad . \tag{3}$$

---

[2)] The Gupta of QED

We found that in this scheme, a top quark of mass as high as 200 GeV could easily be accommodated. In fact we entertained a specific value of 180 GeV for the top quark mass as allowed by the then constraints on the mixing angles describing quark flavor mixing. This value happened to be almost twice the upper bound for the top quark mass in the Fritzsch scheme. The work was presented as a Wayne State University preprint [1] and, of course, was submitted for publication. That manuscript is reproduced here in its entirety [1]. It still bears the manuscript number and the received date assigned to us by the publication office of the journal.[3)] Regretably, due to the author's (SR) departure from Wayne State University to take up permanent position at California State University, Long Beach, the manuscript resulted in two separate publications [2], one in Physical Review D and another in Modern Physics Letters A.

It is to be noted that our primary concern at the time was just to accommodate a heavy top quark and not present a detailed phenomenological analysis. Today, it is amusing to note that the top quark with a mass in the range entertained in the manuscript [1] was discovered [3, 4] in 1995!

### Other related Works

The work on the new scheme of Quark mass matrices was presented by the author at various conferences [5, 6, 7].

The orthogonal matrix **O** that diagonalises **M** was presented in our manuscript in approximate form[4)] in terms of mass ratios $m_i/m_j < 1$. The exact form of **O** is

$$\mathbf{O} = \begin{pmatrix} \sqrt{\frac{(C-m_1)m_2m_3}{C(C-m_2)(C-m_3)}} & \sqrt{\frac{(C-m_2)m_1m_3}{C(C-m_1)(C-m_3)}} & \sqrt{\frac{(C-m_3)m_1m_2}{C(C-m_1)(C-m_2)}} \\ \sqrt{-\frac{(C-m_1)m_1}{(m_1-m_3)(m_2+m_1)}} & \sqrt{\frac{(C-m_2)m_2}{(m_1+m_2)(m_2+m_3)}} & \sqrt{-\frac{(C-m_3)m_3}{(m_3-m_1)(m_2+m_3)}} \\ -\sqrt{\frac{m_1(C-m_2)(C-m_3)}{C(m_2+m_1)(m_1-m_3)}} & \sqrt{\frac{m_2(C-m_1)C-(m_3)}{C(m_2+m_1)(m_2+m_3)}} & -\sqrt{\frac{m_3(C-m_1)(C-m_2)}{C(m_3-m_1)(m_3+m_2)}} \end{pmatrix} . \quad (4)$$

This result was presented at the conference "Beyond the Standard model-IV" [5] and also at the "Symposium on Flavor-changing Neutral Currents: Present and Future Studies" [6]. The matrix **O** (Eq. 4) appears on the front cover of the latter journal proceeding [6]. Later, our scheme was investigated in the works of Du and Xing [8]. In this regard, Ref. [13] of Kang and Kang [9] is most illuminating in summing up most accurately the situation at that point in time. This scheme also surfaced in the studies of Ramond, Ross and Roberts [10] and Fritzsch and Xing [11]. Since then, many groups [12] have joined in the investigation

[3)] The referee report may be requested from the author.

[4)] Eq. (8) of the Wayne State University manuscript.

and vindicated the elegance of the scheme [1] for addressing the problem of flavor mixing. It is to be noted that the matrix **O** in Eq. 4 has found its way in a multitude of forms in the recent literature. Since

$$C + D = m_1 + m_2 + m_3 \quad , \tag{5}$$

all multitude forms are equivalent to the **O** of Ref. [5].

*In all the papers following Refs. [1, 2], the original works have somehow been overlooked.*

## Neutrinos

At the time when the new scheme was conceived, flavor mixing in the neutrino was far from settled. Today the situation is much better. We know that neutrino oscillations require two large mixing angles, while the third angle is tied up with CP violation [13, 14], and is small with no information on the CP violation parameter at the present time. The present experimental information on the mixing angles and the mass squared differences $\Delta m^2_{ij}$ with errors at $2\sigma$ level can be summarized as follows [15, 16]

$$\begin{aligned} &\sin^2\theta_{12} = 0.314(1^{+0.18}_{-0.15}),\ \ \sin^2\theta_{23} = 0.45(1^{+0.35}_{-0.20}),\ \ \sin^2\theta_{13} = (0.8^{+2.3}_{-0.8}) \times 10^{-2}, \\ &\Delta m^2_{21} = 7.92(1 \pm 0.09) \times 10^{-5}\mathrm{eV}^2,\ \ |\Delta m^2_{32}| = 2.6(1^{+0.14}_{-0.15}) \times 10^{-3}\mathrm{eV}^2. \end{aligned} \tag{6}$$

It will be some time before the question of CP viloation in the neutrino sector will be settled.

As it turns out, the Gupta Rajpoot form of mass matrices [1, 2] are also adequate to describe even neutrino flavor mixing [17, 18, 19] modulo the issues shrouding absolute neutrino masses and the $U_{e3}$ element. The scheme in Eqs. (2) and (3) handles both Majorana and Dirac neutrinos. The two additional phases present in the case of Majorana neutrinos are irrelevant for neutrino oscillations.

## A Salam Story and a Call for References to Prior Work[5)]

Appropriate referencing to original literature has always been an issue. My mentor, the late Professor Abdus Salam used to express deep reservations on the subject. In his husky voice, he would say "In our field the one (a person or a group) who gets the most credit is the one who contributes the last, in the appropriate time frame that is". This would then be followed by his loud characteristic laughter that filled the corridors of the building and could be heard as far as the outside of Blackett Laboratories. How right! Judging the situation at hand, today the practice seems to have become the norm.

---

5) For another Salam story, see Ref. [6].

To conclude, understanding the origins of flavor mixing is an issue of transcendential importance in the standard model. Through such studies we may ultimately acquire the knowledge governing fermion masses, flavor mixing and the phenomenon of CP violation. In this regard studies on texture 2(4) zero mass matrices are the initial but important steps. The synopses presented here, it is hoped, has clarified the chronology in the study of texture 2(4) zero mass matrices. Our expectations are that the Gupta Rajpoot scheme [1, 2] will be ovelooked no longer and references to this work will become the norm in all studies on the subject. It is high time that the record be set straight.

# Quark Mass Matrices and the Top-Quark Mass

Suraj N. Gupta and Subhash Rajpoot

Department of Physics, Wayne State University, Detroit, Michigan 48202

If the top quark mass turns out to be greater than about 90 GeV, the Fritzsch ansatz for the mass matrices would run into conflict with existing data on the weak interaction mixing angles. We propose a modified form of mass matrices, in which the diagonal elements corresponding to only the light quarks are vanishingly small. The new scheme is simple to analyze, and more importantly it gives nearly twice the Fritzsch value for the upper bound on the top quark mass when compared with the present knowledge on the Cabibbo-Kobayashi-Maskawa mixing angles.

The quark sector of the standard model contains ten arbitrary parameters which consist of six quark masses, three Cabibbo-like angles, and a CP violating phase. The arbitrariness is reduced to some extent if the patterns of the mixing angles and the phase are related to the masses. Of all the suggested schemes, the scheme of Fritzsch is the most economical[1,2]. However, when the Kobayashi-Maskawa mixing matrix obtained from Fritzsch's form of mass matrices is compared with the experimental results[3], the top quark mass is required to have an upper bound[4] of about 90 GeV. Since the existing data[5] from the colliders puts $m_t > 89$ GeV, the Fritzsch scheme appears to be in trouble. We here propose a scheme based on a modified form of mass matrices.

For the three families of quarks, our mass matrices are of the form

$$M = \begin{pmatrix} o & Ae^{i\alpha} & o \\ Ae^{-i\alpha} & Y & Be^{i\beta} \\ o & Be^{-i\beta} & C \end{pmatrix}, \quad Y \ll C, \tag{1}$$

where the choice of the diagonal elements is inspired by the mass hierarchy of the quarks (u, c, t) and (d, s, b). For Y = o, (1) reduces to the Fritzsch mass matrix.

The matrix M is expressible as

$$M = P\overline{M}P^{\dagger}, \tag{2}$$

where

$$\overline{M} = \begin{pmatrix} o & A & o \\ A & Y & B \\ o & B & C \end{pmatrix}, \quad P = \begin{pmatrix} 1 & o & o \\ o & e^{-i\alpha} & o \\ o & o & e^{-i(\alpha+\beta)} \end{pmatrix}, \tag{3}$$

and $\overline{M}$ can be diagonalized by means of the transformation

$$\overline{M} = OM_{diag}O^{\dagger} \tag{4}$$

with

$$M_{diag} = \begin{pmatrix} m_1 & o & o \\ o & -m_2 & o \\ o & o & m_3 \end{pmatrix}. \qquad (5)$$

It then follows that

$$\begin{aligned} & m_1 - m_2 + m_3 = C + Y, \\ & m_1m_3 - m_1m_2 - m_2m_3 = CY - A^2 - B^2, \qquad (6) \\ & m_1m_2m_3 = A^2C, \end{aligned}$$

or

$$A = \left( \frac{m_1m_2m_3}{m_1 - m_2 + m_3 - Y} \right)^{\frac{1}{2}},$$

$$B = \left( - \frac{m_1m_2m_3}{m_1 - m_2 + m_3 - Y} + CY + m_1m_2 + m_2m_3 - m_1m_3 \right)^{\frac{1}{2}}, \qquad (7)$$

$$C = m_1 - m_2 + m_3 - Y.$$

Upon retaining terms in leading powers of $\frac{m_1}{m_2}$, $\frac{m_1}{m_3}$, $\frac{m_2}{m_3}$, $\frac{Y}{m_3}$ and in view of our knowledge of the quark masses[6] , O is found to be

$$O = \begin{pmatrix} 1 & -\sqrt{\frac{m_1}{m_2}} & \sqrt{\frac{m_1}{m_3}}\,\frac{\sqrt{m_2(m_2+Y)}}{m_3} \\ \sqrt{\frac{m_1}{m_2}} & 1 & \sqrt{\frac{m_2+Y}{m_3}} \\ -\sqrt{\frac{m_1(m_2+Y)}{m_2 m_3}} & -\sqrt{\frac{m_2+Y}{m_3}} & 1 \end{pmatrix}. \tag{8}$$

According to (2) and (4),

$$M = RM_{diag}R^{\dagger} \quad \text{with} \quad R = PO, \tag{9}$$

which yields the K-M matrix

$$V = R^{u\dagger}R^{d} = O^{u\dagger}P^{ud}O^{d}, \tag{10}$$

where

$$P^{ud} = P^{u\dagger}P^{d} = \begin{pmatrix} 1 & o & o \\ o & e^{i\phi_1} & o \\ o & o & e^{i\phi_2} \end{pmatrix}, \tag{11}$$

and

$$\phi_1 = \alpha^u - \alpha^d,\ \phi_2 - \phi_1 = \beta^u - \beta^d . \tag{12}$$

The matrix elements of V, obtained by the substitution of (8) and (11) into (10), are

$$V_{ud} = 1, \tag{13}$$

$$V_{us} = -\sqrt{\frac{m_d}{m_s}} + e^{i\phi_1}\sqrt{\frac{m_u}{m_c}}, \tag{14}$$

$$V_{ub} = \sqrt{\frac{m_d}{m_b}}\ \frac{\sqrt{m_s\,(m_s + Y^d)}}{m_b} + e^{i\phi_1}\sqrt{\frac{m_u}{m_c}}\left(\frac{m_s + Y^d}{m_b}\right)^{\frac{1}{2}} - e^{i\phi_2}\sqrt{\frac{m_u}{m_c}}\left(\frac{m_c + Y^u}{m_t}\right)^{\frac{1}{2}}, \tag{15}$$

$$V_{cd} = -\sqrt{\frac{m_u}{m_c}} + e^{i\phi_1}\sqrt{\frac{m_d}{m_s}}\ , \tag{16}$$

$$V_{cs} = e^{i\phi_1}, \tag{17}$$

$$V_{cb} = e^{i\phi_1}\left(\frac{m_s + Y^d}{m_b}\right)^{\frac{1}{2}} - e^{i\phi_2}\left(\frac{m_c + Y^u}{m_t}\right)^{\frac{1}{2}}, \tag{18}$$

$$V_{td} = \sqrt{\frac{m_u}{m_t}}\ \frac{\sqrt{m_c(m_c + Y^u)}}{m_t} + e^{i\phi_1}\sqrt{\frac{m_d}{m_s}}\left(\frac{m_c + Y^u}{m_t}\right)^{\frac{1}{2}} - e^{i\phi_2}\sqrt{\frac{m_d}{m_s}}\left(\frac{m_s + Y^d}{m_b}\right)^{\frac{1}{2}}, \tag{19}$$

$$V_{ts} = e^{i\phi_1}\left(\frac{m_c + Y^u}{m_t}\right)^{\frac{1}{2}} - e^{i\phi_2}\left(\frac{m_s + Y^d}{m_b}\right)^{\frac{1}{2}}, \tag{20}$$

$$V_{tb} = e^{i\phi_2}\ . \tag{21}$$

The above elements give the weak interaction mixing matrix in terms of the quark masses. In order to relate the elements of V to experiments, we take the standard parametrization of the K-M matrix[3]

$$V = \begin{pmatrix} c_{12}c_{13} & c_{13}s_{12} & s_{13}e^{-i\delta} \\ -c_{23}s_{12}-c_{12}s_{23}s_{13}e^{i\delta} & c_{12}c_{23}-s_{12}s_{23}s_{13}e^{i\delta} & c_{13}s_{23} \\ s_{12}s_{23}-c_{12}c_{23}s_{13}e^{i\delta} & -c_{12}s_{23}-s_{12}c_{23}s_{13}e^{i\delta} & c_{23}c_{13} \end{pmatrix}, \tag{22}$$

where $c_{ij} = \cos\theta_{ij}$ and $s_{ij} = \sin\theta_{ij}$, and $\theta_{12}$ is essentially the Cabibbo angle with the experimental value $s_{12} = 0.22$.

The present data[3] from semileptonic decays of the B mesons and charmless B decays together with the unitarity of the V matrix requires $|S_{23}| \leq 0.053$ and $|S_{13}|/|S_{23}| \leq 0.013$. In view of these constraints, we can set $c_{13} \simeq 1$ and $c_{23} \simeq 1$, and by comparing (14), (15), and (18) with (22), arrive at

$$s_{12} = -\sqrt{\frac{m_d}{m_s}} + e^{i\phi_1}\sqrt{\frac{m_u}{m_c}}, \tag{23}$$

$$s_{13} = e^{i\delta}\sqrt{\frac{m_u}{m_c}}\left[e^{i\phi_1}\left(\frac{m_s + Y^d}{m_b}\right)^{\frac{1}{2}} - e^{i\phi_2}\left(\frac{m_c + Y^u}{m_t}\right)^{\frac{1}{2}}\right] + e^{i\delta}\sqrt{\frac{m_d}{m_b}}\,\frac{\sqrt{m_s(m_s + Y^d)}}{m_b}, \tag{24}$$

$$s_{23} = e^{i\phi_1}\left(\frac{m_s + Y^d}{m_b}\right)^{\frac{1}{2}} - e^{i\phi_2}\left(\frac{m_c + Y^u}{m_t}\right)^{\frac{1}{2}}. \tag{25}$$

We find that neither $s_{12}$ nor $|s_{13}|/|s_{23}|$ is significantly altered by our modification of the mass matrices. But, it now follows from (25) that

$$|s_{23}| \geq \left(\frac{m_s + Y^d}{m_b}\right)^{\frac{1}{2}} - \left(\frac{m_c + Y^u}{m_t}\right)^{\frac{1}{2}}, \tag{26}$$

so that the upper bound for the top quark mass is given by

$$m_t \leq \frac{m_c + Y^u}{\left(\sqrt{\frac{m_s + Y^d}{m_b}} - |s_{23}|\right)^2}. \tag{27}$$

Since the diagonal elements in our mass matrix (1) are inspired by the hierarchy of the quark masses and, according to (7), C is somewhat smaller than $m_3$, it is reasonable to expect that $Y < m_2$. We further note that $Y^u$ increases the upper bound for the top quark mass, while $Y^d$ decreases it.

A special form of our scheme is a hybrid of the Fritzsch scheme and our scheme which is gotten by taking $Y^d=0$ .If we assume further that $Y^d \approx m_c$ , then the upper bound on $m_t$ is

$$m_t \leq \frac{2m_c}{\left(\sqrt{\frac{m_s}{m_b}} - |s_{23}|\right)^2} , \qquad (28)$$

which is twice the result obtained from the Fritzsch scheme[4].

Clearly the scheme proposed here can easily accommodate $m_t$ as high as the electroweak scale. The main point of our analysis is that if the top quark turns out to be very heavy then some of the zero elements of the Fritzsch quark mass matrices may not be zero. In particular , very heavy top mass can easily be accommodated if the quark mass matrices have only one zero diagonal element and non vanishing nearest neighbour interactions as in (1) . CP violation and related issues in this new scheme will be presented elsewhere.